\documentclass[twocolumn]{article}
\usepackage{graphicx} 
\usepackage{amsmath}
\usepackage{geometry} % <-- Add this package
\usepackage{subcaption}
\usepackage{cleveref}
\usepackage{authblk} % for affiliations
\title{Geometry-Driven Charge and Spin Transport in $\beta 12$ Borophene Quantum Dots}
\author[1]{Seyed Mahdi Mastoor\thanks{smahdimastoor@gmail.com}}
\author[1]{Amirhossein Ahmadkhan Kordbacheh\thanks{akordbacheh@iust.ac.ir}}

\affil[1]{Department of Physics, Iran University of Science and Technology, Tehran, Iran}

\date{} % leave empty to omit date

\DeclareUnicodeCharacter{2212}{-}
\begin{document}

\maketitle
\begin{abstract}

Theoretical research has been conducted to study how geometry affects charge and spin transport in $\beta 12$ borophene quantum dots, which are confined systems. The study examined two distinct central regions, which included a circular disc and a regular hexagonal area that connected to semi-infinite zigzag and armchair borophene nanoribbon leads. The system was described by a five-band tight-binding Hamiltonian parameterized using first-principles data, and the transport properties were calculated within the non-equilibrium Green’s function framework.Spin resolved transmissions and spin polarization were computed for a range of lead widths and proximity-induced exchange field strengths. The analysis revealed distinct transport characteristics determined by geometry and edge configuration: armchair-connected structures exhibited broader and more stable fully spin-polarized windows compared with zigzag-connected counterparts. Furthermore, critical lead-width thresholds ($ \approx1.01$ nm for zigzag and $\approx0.87$ nm for armchair) and a moderate exchange field above which complete spin filtering occurs were identified. The results highlight the strong influence of edge termination and confinement geometry on transport properties and provide useful design guidelines for developing borophene-based nanoscale spintronic devices.
\end{abstract}

\section{Introduction}
The field of nanoelectronics alongside spintronics has found a new frontier through two-dimensional materials, which demonstrate reduced dimensionality and tunable electronic structure, along with strong edge and geometry sensitivity. The atomic layers of 2D materials enable control of charge and spin transport for developing different devices that combine low-power transistors, spin filters, and quantum information components \cite{Jia2025-mz} \cite{Zeng2024-xe}. Research about 2D van der Waals systems including their heterostructures proves that magnetic proximity effects cause non-magnetic layers to develop significant exchange splittings (such as graphene–CrSBr). The technique allows spin polarization to be controlled through electric fields \cite{Yang2024-lg}. The scientific community has achieved experimental success in developing 2D ferromagnets and topological quantum materials such as VC$_6 $and NbC$_6$ monolayers that exhibit natural magnetic behavior while demonstrating unique transport properties including the quantum anomalous Hall effect. Recent evaluations have been conducted to assess the potential benefits and implementation challenges associated with developing stable two-dimensional spin-polarized quantum materials for practical applications. These include room temperature conditions and irregular edge structures \cite{Zhang2025-fn}\cite{Wu2024-pu}.

Recently, another two-dimensional substance called borophene has been synthesized on a silver substrate through ultra-high-vacuum conditions according to experimental results which matched previous theoretical predictions \cite{Mannix2015-uj} \cite{Feng2016-mu}. The polymorph $\beta 12$ borophene stands out because it combines fast carrier movement with strong directional differences, and its distinctive transition between metallic and semiconducting properties. The research demonstrates that bilayer borophene exhibits semiconducting behavior while maintaining exceptional carrier mobility values that exceed$ 2×10^4 \space cm^2 V^{-1} s^{-1}$ \cite{Yan2023-rn}. The combination of borophene with graphene, or boron-carbide layers in heterostructures enables adjustable band alignment and shows excellent optoelectronic characteristics\cite{Niu2024-zj}. The research on borophene for flexible electronics and quantum transport applications demonstrates that its charge and spin transport characteristics depend on geometry (shape, edge type, and confinement), and external fields \cite{Nikan2022-sj} \cite{Nikan2024-hu}.
$\beta 12 $ borophene distinguishes itself through several recent findings that underscore its promise for electronic and optoelectronic applications. The first-principles studies showed that multilayer $\beta 12 $ borophene preserves its anisotropic metallic characteristics while increasing the number of layers enhances both interlayer charge transfer and band splitting. This occurs more strongly in $\beta 12 $ than in other allotropes \cite{Wang2024-xl}. The bilayer $\beta 12 $ like phase grown on Cu(111) exhibits two primary characteristics which include strong metallic conductivity, and enhanced thermoelectric, and optical absorption across visible and infrared wavelengths. These properties demonstrate effective interaction between the substrate and borophene layers \cite{Ali2024-mr}. Rectangular $\beta 12 $ quantum dots show particular energy level degeneracies and optical absorption changes based on their size when they exist in nanoscale confined spaces: The width expansion leads to blueshifts in particular peaks, but the length creates redshifts in different peaks; The degeneracies between sets of levels (E$_1$, E$_2$) that depend on width demonstrate how quantum confinement and edge geometry affect these levels \cite{Liang2025-xt}. Moreover, phenomena such as Klein tunneling have been reported in $\beta 12 $  borophene, revealing unusually perpendicular transport behaviors, since charge carriers can transmit through potential barriers with high probabilities under certain orientations \cite{Lai2024-pj}. The effect of lattice vibrations / phonons on spin and charge transport in $\beta 12 $ nanostructures has been recently analyzed: investigations into nanoribbons show that electron-phonon coupling (EPC) and structural edge type (zigzag vs armchair) significantly modulate spin-resolved current and can either suppress or enhance transmission depending on edge orientation, thus offering a further tuning parameter for device design \cite{Davoudiniya2025-wl}.These results demonstrate The findings demonstrate that $\beta 12 $ borophene's electronic characteristics show strong dependence on the number of layers, substrate contact, edge configuration, and spatial restrictions.

Although there has been substantial progress in understanding the transport and electronic properties of $\beta 12 $ borophene,  yet essential details about its confined geometries and spin-resolved properties remain unexplored. Many works have studied nanoribbons of $\beta 12 $ borophene, analyzing the effects of edge type (zigzag vs armchair), strain, electric fields, or impurities on transmission probabilities, current–voltage curves, and metal-to-semiconductor transitions \cite{Davoudiniya2021-bl}\cite{Norouzi2021-qi}. However, these ribbon studies assume quasi-one-dimensional geometry and semi-infinite leads; they do not adequately capture the behavior of quantum dot or finite geometries (e.g. discs or hexagons), where confinement in more than one spatial direction, finite size effects, and lead coupling can produce qualitatively different transmission spectra. The current research lacks a systematic evaluation of how shape elements (disc and hexagon), lead geometry (width and orientation), and exchange field strength affect charge and spin transmission in finite geometries.

The main goal of this research is to develop specific design guidelines which establish relationships between quantum dot shape (disc versus hexagon), size (effective radius), and edge termination (zigzag, versus armchair leads) with their effects on charge and spin transport in confined $\beta 12 $ borophene quantum dots. We employ a tight-binding model calibrated to first-principles calculations and compute transport via the non-equilibrium Green’s function formalism. The central scatterer (disc or hexagon) is connected to semi-infinite leads, and spin-resolved retarded/advanced Green’s functions are used to obtain T(E) and and the spin polarization P(E). We conduct organized examinations of lead width and proximity-induced exchange field M to determine how zero-transmission zones, transmission plateaus, and fully spin-polarized windows change with different geometric configurations and contact designs. The generated maps serve three main objectives, which include (i) finding parameter ranges that achieve maximum spin filtering performance while maintaining suitable conductance levels, and (ii) assessing how these ranges react to changes in M and lead matching. This delivers experimental guidelines for building borophene-based nanoscale spintronic devices.

The remainder of this paper is organized as follows: Section 2 describes the tight-binding Hamiltonian and the NEGF formalism used for transport calculations. The third section includes the findings from the study, which involve both circular and hexagonal shapes. Section 4 summarizes the main conclusions and potential applications in spintronic device design.

\section{Theoretical Model}

\subsection{Tight-binding Hamiltonian}
This research employs a five-band tight-binding (TB) model to analyze the electronic properties of monolayer $\beta_{12}$-borophene around $K$ and $K^\prime$ points in the Brillouin zone. The lattice structure of $\beta_{12}$-borophene consists of two lattice constants $2.92 \AA$   and $5.06 \AA$ while the TB Hamiltonian hopping parameters are obtained through matching its dispersion with first-principles calculations \cite{Ezawa2017-lw}. The $P_z$ orbitals in the boron atoms represent the only basis functions in this model but previous ab initio calculations demonstrate its accuracy for describing the fundamental low-energy characteristics of $\beta_{12}$-borophene because of its almost flat structure \cite{Feng2017-ge}. Among the three available TB models (homogeneous, inversion symmetric (IS), and  inversion non-symmetric (INS)) the INS model provides the most accurate description since the Ag substrate interaction removes inversion symmetry and produces a small Dirac spectrum gap and INS is experimentally closer
to its real structure\cite{Ezawa2017-lw} \cite{Feng2017-ge}.

The charge carriers in pristine borophene do not possess spin polarization by default but insulating ferromagnetic substrates can induce magnetic characteristics into borophene leads through proximity \cite{Bishnoi2013-kw}. The exchange field from this substrate functions like an external magnetic field which disrupts time-reversal and spin symmetry inside the system.

\begin{figure}
    \centering
    \includegraphics[width=1\linewidth]{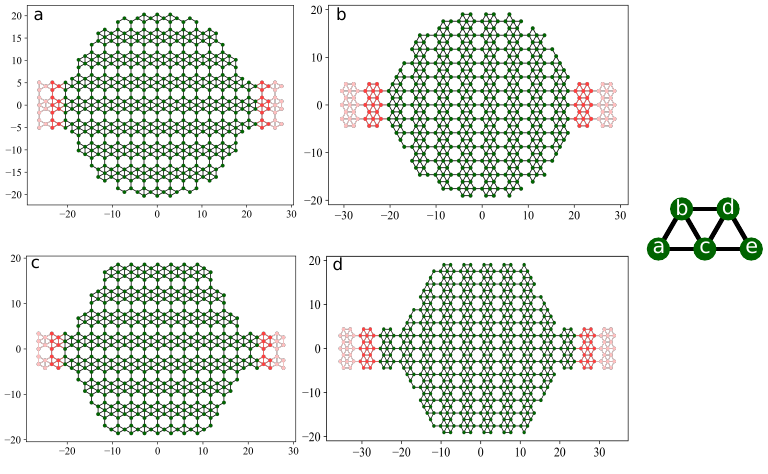}
    \caption{Schematic illustration of the studied structures: disc geometries with (a) zigzag and (b) armchair edge orientations, and hexagonal geometries with (c) zigzag and (d) armchair edge orientations. The unit cell of $\beta_{12}$ borophene is shown on the left side of the figure for reference.  }
    \label{shematic of disc}
\end{figure}

The tight-binding (TB) Hamiltonian of $\beta_{12}$-borophene in the presence of exchange  fields can be expressed as:
\begin{equation}
\begin{split}
    H = & \sum_i \varepsilon_{i\sigma} C_{i\sigma}C_{i\sigma}^\dagger 
    + 
    \sum_{<ij>\sigma\sigma^\prime}
    t_{ij}^{\sigma\sigma^\prime}
    C_{j\sigma^\prime}C_{i\sigma}^\dagger \\
    & + M\sum_i C_{i\sigma} \sigma_z C_{i\sigma}^\dagger 
\end{split}
\end{equation}

The first term represents the onsite energies, where the values of $\varepsilon_i$ are assigned based on the unit cell configuration illustrated in Fig.~\ref{shematic of disc}. For the INS model \cite{Ezawa2017-lw}, these parameters are defined as $\varepsilon_a = \varepsilon_d = 0.196$ eV, $\varepsilon_b = \varepsilon_e = -0.058$ eV, and $\varepsilon_c = -0.845$ eV. The second term corresponds to the hopping interactions between nearest-neighbor atoms, with hopping parameters given by $t_{ab} = t_{de} = -2.04$ eV, $t_{ac} = t_{ce} = -1.79$ eV, $t_{bc} = t_{cd} = -1.84$ eV, $t_{ae} = -2.12$ eV, and $t_{bd} = -1.97$ eV. The third term, which contains the z-component of the Pauli matrix $\sigma_z$, describes the exchange magnetic field of strength $M$ that is induced through the proximity effect from an adjacent ferromagnetic substrate. Also, the $C_{j\sigma^\prime}C_{i\sigma}^\dagger $operator creates (annihilates) an electron at site $i$ with spin $\sigma$ and$<i,j>$denotes the sum over
sites.
The disc geometry is defined by the condition $x^2+y^2\leq R^2 $
, while the hexagonal geometry is constructed based on the intersection of six linear boundaries forming a regular hexagon with circumradius 
R. Here, R denotes the effective radius measured from the center of the structure to its vertices, determining the overall size of the scattering region.

Since the leads can be regarded as a periodic sequence of unit cells along the $x$ direction, its Hamiltonian, according to Bloch’s theorem, can be expressed as a function of the wave vector $k_x$ as follows:
\begin{equation}
    H(k_x)=H_te^{ik_xa}+H_0+H_te^{-ik_xa}
\end{equation}
Furthermore, diagonalizing this Hamiltonian yields the corresponding band structure and energy dispersion. Here, 
$H_0$ represents the Hamiltonian of an individual unit cell within the lead, $H_t$denotes the hopping interaction between neighboring unit cells, and 
$a$ is the lattice constant.
\subsection{Spin-dependent quantum conductance}
Our model connects the central scattering region to two semi-infinite borophene nanoribbon (BNR) leads through either a disc or hexagon configuration. The iterative scheme from Sancho et al. \cite{Teichert2019-er} enables fast and steady convergence for periodic systems to calculate surface Green’s functions of semi-infinite leads, which serve as transport property evaluation tools. The described technique enables quick computation of lead self-energies and their connection to the finite central region. The transport calculations operate under ballistic conditions. At low temperatures, electrons are assumed to move without scattering and without phonon interactions \cite{Giustino2017-gf}. In this case, one
can apply the Landauer-Büttiker formula for the spin dependent transmission as \cite{Datta2012-ig}:
\begin{equation}
    T^s(E)=Trace[\Gamma_L^s(E)G^s(E)\Gamma_2^s(E) G^s(E)^\dagger]
\end{equation}
where $T^s$ indicates the transmission from left to right leads for the carrier with spin $s$. $G^s(E) , G^s(E)^\dagger$ are retarded (advanced) Green’s function and defined by :
\begin{equation}
 G^s(E) = G^s(E)^\dagger=[E-H-\Sigma^{R,A}_L-\Sigma^{R,A}_R]^{-1}
\end{equation}
The broadening matrix $\Gamma(E)$ due to strong couplings between the central region and left (right) lead of the system, where define as :
\begin{equation}
    \Gamma_{L,R}^s(E) = i(\Sigma_{L,R}^s - (\Sigma_{L,R}^s)^\dagger )
\end{equation}
Spin dependent self-energy terms,$\Sigma_{L,R}^s $, corresponding to the left and right leads, can be numerically evaluated using an iterative recursive algorithm \cite{Sancho1984-dp}. Note that the totat transmission of system obtained as $T(E)=T^{\uparrow}(E) + T^{\downarrow}(E)$. 

The spin polarization, defined in terms of the spin-resolved conductance, is expressed as:
\begin{equation}
    P_s(E)=\dfrac{T^{\uparrow}(E) - T^{\downarrow}(E)}{T^{\uparrow}(E) + T^{\downarrow}(E)}
\end{equation}
It is important to note that the spin polarization $P_S$ varies within the range $-1\leq P_S \leq +1$. A positive value of $P_S$ indicates that spin-up electrons dominate the transport process, whereas a negative value $P_S$ signifies that spin-down electrons contribute more strongly to the overall conductance. The magnitude of 
$P_S$ reflects the degree of spin imbalance in the transmitted current—values close to $\pm 1$ correspond to nearly complete spin polarization, while those near zero represent unpolarized or weakly polarized transport. The system performance as a spin filter operates at a fundamental level through this parameter. It enables spintronic devices to function properly \cite{Leitao2024-jl}.

\section{Results and Discussion }

\subsection{Disc}
The disc radius for the zigzag setup was set at 4.05 nm and the lead width was 1.01 nm while the armchair setup used a 3.80 nm radius with a 0.87 nm lead width. These dimension values were selected to maintain the basic disc shape with soft edges for systematic investigation of both zigzag and armchair terminations. These specific size parameters match typical theoretical and experimental studies which show nanostructures of a few nanometers demonstrate strong quantum confinement and edge-related phenomena \cite{Bazrafshan2023-cm} \cite{Jafari2022-yw}. The lead widths chosen in this study allow for proper nanodisc connection along with stable propagating modes which lead to a valid analysis of how different edge types affect charge and spin transport.

\begin{figure}[h!]
    \centering
    \includegraphics[width=1\linewidth]{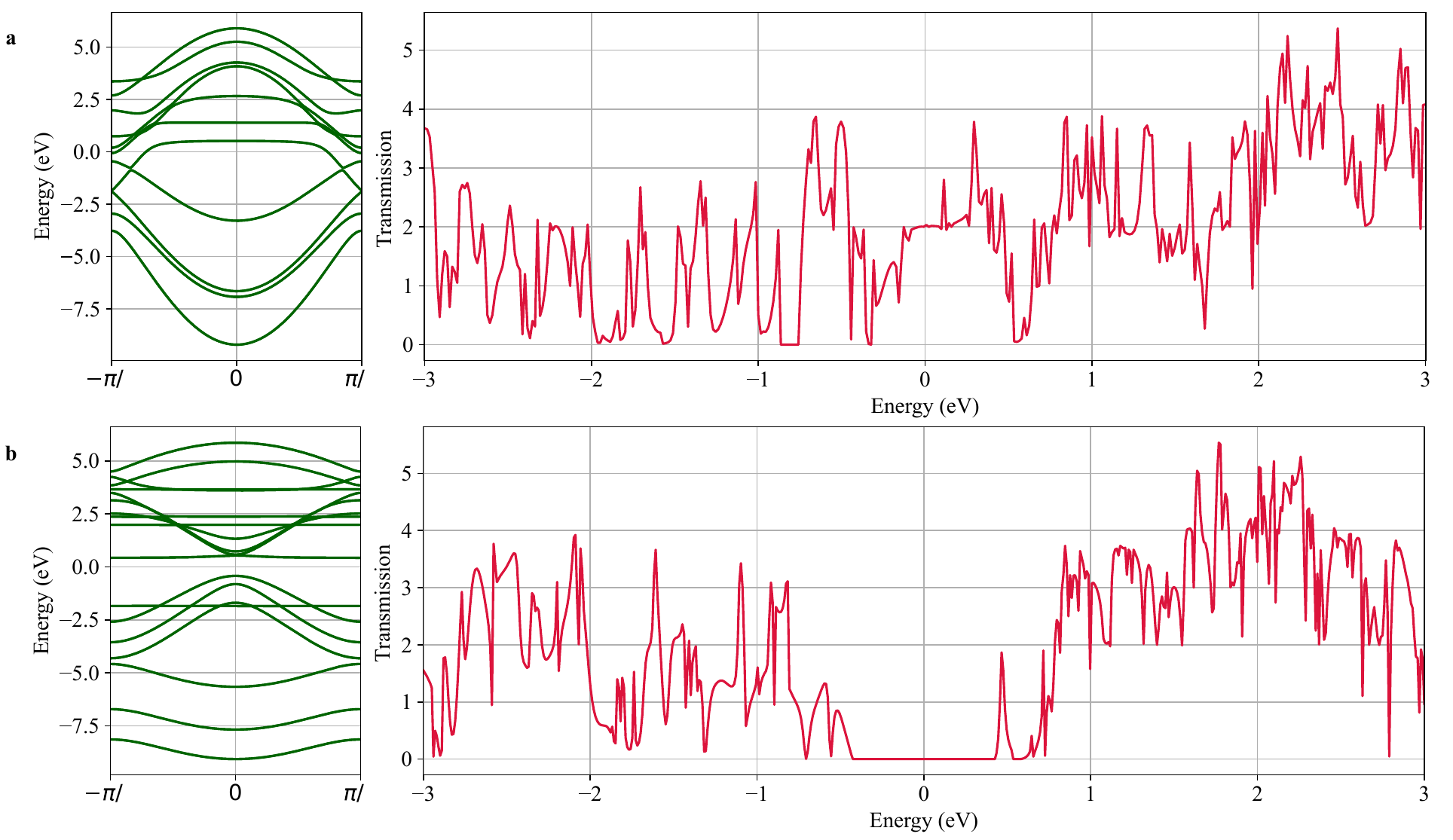}
    \caption{Band structure (left ) and charge transmission (right ) for borophene nanodisc systems with edge orientations of (a) zigzag and (b) armchair. }
    \label{T_D}
\end{figure}

Fig.~\ref{T_D}(a) shows gapless behavior in the band structure of the disc with zigzag edge leads. On the other hand, the corresponding charge transmission shows multiple points of zero transmission along with a brief plateau. The change from propagating to non-propagating charge transport modes is reflected in the observable fluctuations caused by these zero-transmission points.
The band structure reveals available states in both the leads and central region, but does not show how well these states connect with each other. The BNR as a lead contains metallic states near the Fermi level, but quantum interference at the lead–disc interfaces blocks transmission. The curved shape of the disc, together with multiple scattering routes, causes propagating modes to cancel out their phases, which produces destructive interference that blocks transmission at specific energy levels. The zero-conductance points in the device occur because of poor device coupling and interference effects that create localization even though states remain accessible \cite{Thoss2018-xm}.
Fig.~\ref{T_D}(b) illustrates the presence of a flat band near the Fermi energy, along with the opening of a band gap for the armchair edge orientation where is in. The corresponding transmission spectrum reveals a relatively long plateau, which is consistent with the features observed in the band structure. Additionally, isolated zero-transmission points can be identified, indicating localized interruptions in charge transport.  
\begin{figure}
    \centering
    \includegraphics[width=1\linewidth]{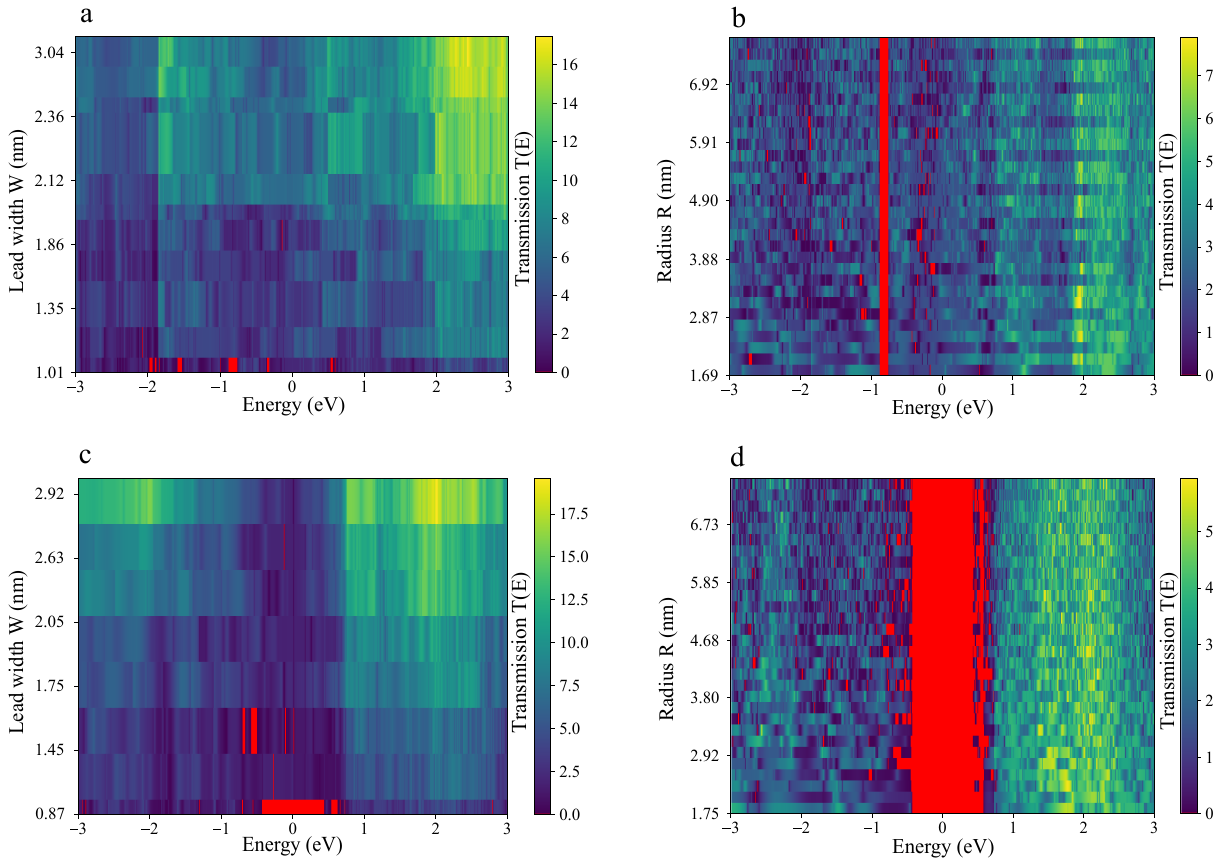}
    \caption{Transmission of initial desired disc with zigzag edges (upper row) and armchair edges (bottom row) by changing leads width (a, c) and disc radius (b, d). Red color indicate zero transmission.}
    \label{heat map of r ,w of disc}
\end{figure}

When the leads width is increased, the transmission spectra display additional transport channels, which arise due to the activation of additional propagating charge carrier modes. This can be explained by the expected impact of quantum confinement; wider leads relieve confinement and allow several subbands to be involved in the conduction \cite{Sverdlov2022-sn}. For zigzag edges, zero-transmission points appear only when the lead width is below about 1.01nm, as shown in Fig.~\ref{heat map of r ,w of disc}(a). This suggests that very narrow leads enhance destructive interference and confinement effects, which can block charge propagation entirely. In contrast, when the width is larger, the system supports continuous conducting pathways. For armchair edges Fig.\ref{heat map of r ,w of disc}(c), distinct zero-transmission features are observed below 0.87nm, with additional interruptions in transport appearing around 1.5nm, highlighting the edge-dependent nature of confinement in these systems.
Conversely, varying the disc's radius has little visible effect on zero-conduction regions Fig.\ref{heat map of r ,w of disc}(b,d). The zero-transmission locations are still largely at the same energies. This serves to highlight that the suppression of transport is being controlled mainly by the confinement and the quantum interference in the leads rather than the geometrical radius of the disc itself. Even though special geometry of central region impact on wave propagation in leads and change electronic structure of them. 

\begin{figure}
    \centering
    \includegraphics[width=1\linewidth]{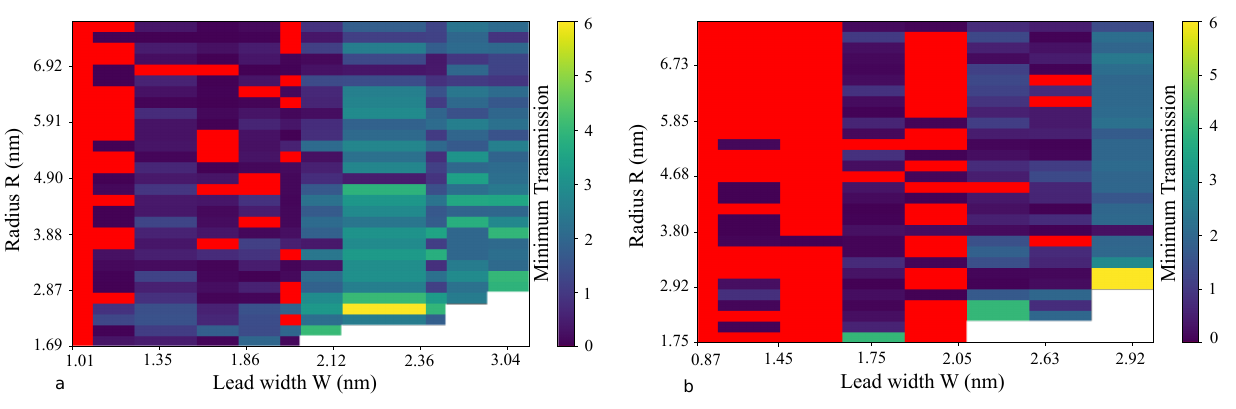}
    \caption{Minimum transmission (red color) per vary range of lead width and disc radius with zigzag (a) and armchair (b) edges. white zone are related to those leads and disc size where has mismatch in simulations.}
    \label{H_r_w_disc}
\end{figure}

Fig.~\ref{H_r_w_disc} shows that, irrespective of the disc radius, both zigzag and armchair edges exhibit zero-transmission  zones at lead widths around 1.01 nm and 
0.87 nm, respectively. For zigzag edges Fig.~\ref{H_r_w_disc}(a), increasing the lead width eventually reaches an upper threshold beyond which zero-transmission no longer persists. This behavior can be attributed to the fact that zigzag borophene nanoribbon (ZBNRs) display metallic characteristics in the delocalized regime, allowing charge carriers to propagate without strong confinement once the width exceeds a critical value \cite{Izadi_Vishkayi2018-il} \cite{Nikan2022-sj}. In contrast, armchair edges Fig.~\ref{H_r_w_disc}(b) which intrinsically exhibit semiconducting behavior below their critical width retain a broader range of zero-transmission regions as both the lead width and disc radius increase. This difference highlights the edge-dependent nature of transport, While zigzag edges tend toward metallic conduction in wider configurations, armchair edges sustain confinement-induced transport suppression over a wider parameter space due to their semiconducting band structure.
\subsubsection{Spin Transport of Disc Geometry}

Borophene nanoribbons are known to support nontrivial edge states protected by time-reversal symmetry (TRS), which strongly influence their transport behavior. Breaking TRS, or equivalently spin-inversion symmetry, can be achieved by introducing an exchange field through the proximity effect of a ferromagnetic insulator placed beneath the nanoribbon. Compared with alternative approaches, such as transition metal doping \cite{Chang2013-eq} \cite{Feng2017-me} or structural buckling \cite{Yan2021-kf}, this method is purely external and short ranged, thereby minimizing any detrimental impact on electron mobility. This makes proximity induced exchange fields a promising route for tailoring spin-dependent transport in borophene-based systems.

\begin{figure}[h]
    \centering
    \includegraphics[width=1\linewidth]{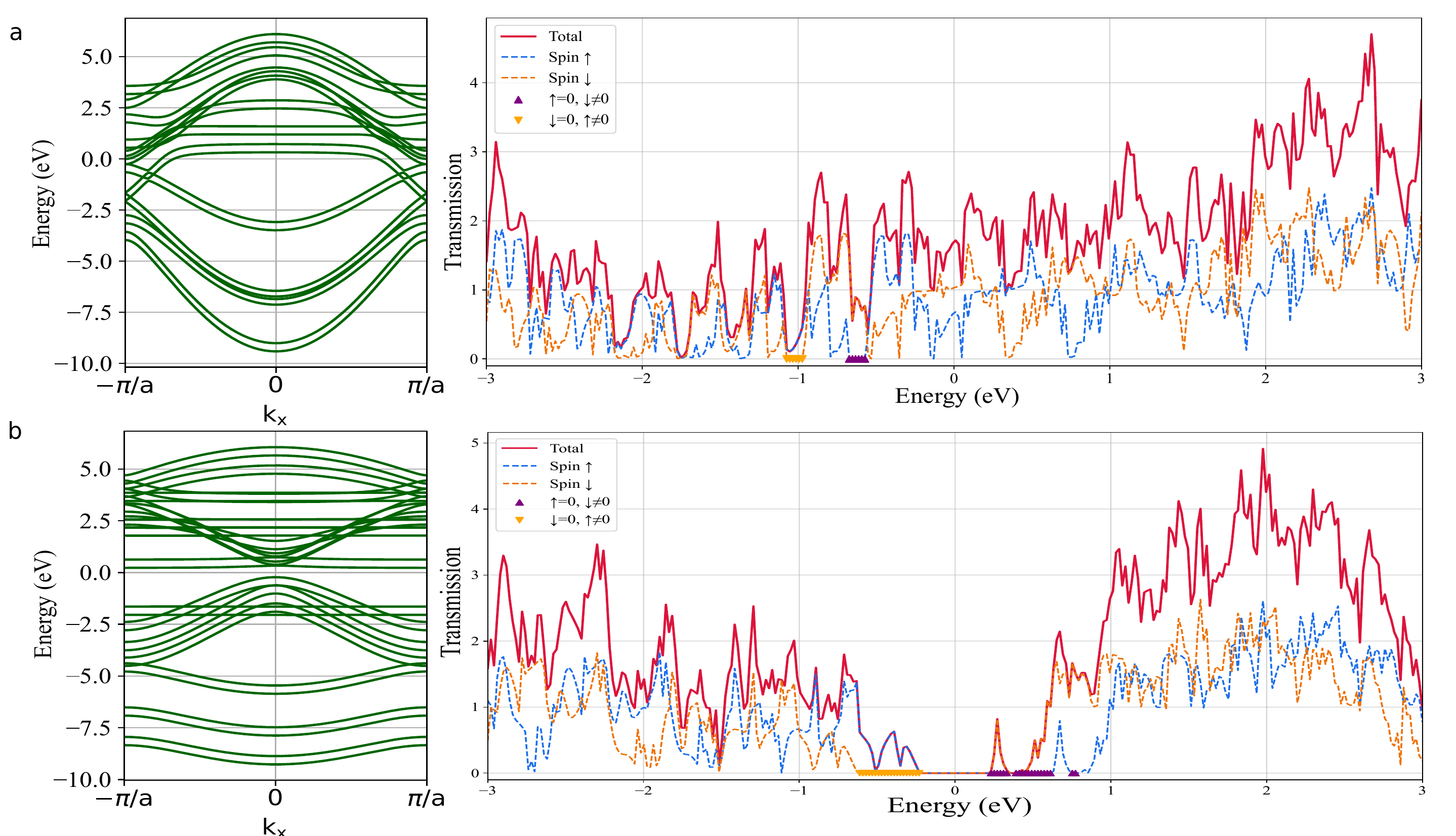}
    \caption{Band structure (left ) and spin-resolved transmission (right ) for borophene nanodisc systems with edge orientations of (a) zigzag and (b) armchair.}
    \label{T_m_d}
\end{figure}

As illustrated in fig.\ref{T_m_d}, by imposing nonlocal exchange field ( M= $0.2$eV), the spin-up and spin-down bands of leads are split and shifted in opposite directions in energy.The splitting occurs because the exchange field breaks the spin degeneracy by creating an effective magnetic interaction which affects spin-up and spin-down electrons differently. The energy levels of the two particles move in opposite directions because of exchange coupling between them \cite{Pei2022-ae}.  Although overall transmission shows non-zero for all range of energy, distinct windows of complete spin polarization emerge. For zigzag edges Fig.\ref{T_m_d}(a), a fully spin-up–polarized current is observed between -1.07 eV and −0.97 eV, while a fully spin-down–polarized current occurs in the range from −0.67 eV to −0.57 eV. In the case of armchair edges Fig.\ref{T_m_d}(b), the spin-polarized regions are noticeably broader, and a zero-transmission zone persists. Specifically, complete spin-up polarization appears between $−0.61$ eV and $-0.23$ eV, whereas spin-down polarization is found in the intervals $0.23$ eV to $0.61$ eV and $0.75$ eV to $0.85$ eV. The findings indicate that edge orientation strongly controls spin-selective transport since armchair edges deliver extended  operational windows for spin filtering. The disc geometry investigated here links the central scattering region to  semi-infinite leads through which a nonlocal exchange field produces strong spin polarization within defined energy intervals.  The polarization intervals stay tight for zigzag-connected leads because zigzag ribbons maintain metallic characteristics during the delocalized  state. The armchair-connected leads demonstrate extended intervals of full spin polarization while maintaining zero transmission regions which  match their semiconducting nature \cite{Norouzi2021-qi}. The combination between disc shape and lead alignment creates an energy-selective spin  filter platform which enables precise control of nanoscale device spin filtering.

\begin{figure}
    \centering
    \includegraphics[width=1\linewidth]{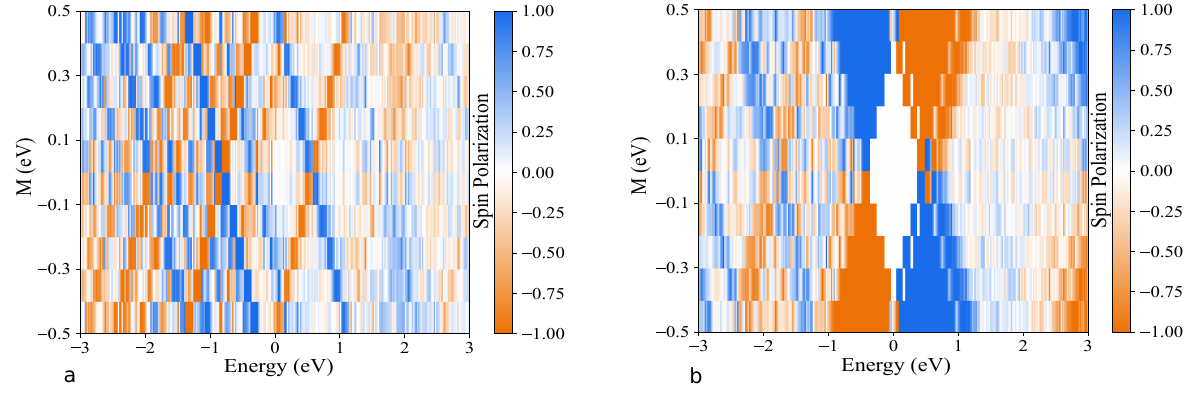}
    \caption{Spin polarization for vary exchange fields for disc geometry with zigzag (a) and armchair (b) edges as function of the energy of incoming electrons.}
    \label{spin_p_d}
\end{figure}

To investigate the role of edge geometry in determining the spin polarization of the borophene nanodisc (BND), we calculated the contour map of the spin polarization, as shown in Fig.~\ref{spin_p_d}, as a function of the exchange field strength $M$ and the energy of the incoming carriers. For zigzag edges Fig.~\ref{spin_p_d}(a), the results reveal only narrow regions of complete spin-up or spin-down polarization, reflecting the metallic-like character of zigzag terminations. In contrast, the disc with armchair edges exhibits well-defined and significantly broader regions of spin polarization, extending across a wide energy window from approximately -1eV to +1eV fig.\ref{spin_p_d}(b).

\subsection{Hexagon}

The hexagonal design for zigzag borophene nanostructure used a 3.70 nm radius with 0.84 nm lead width while the armchair design used a 3.80 nm radius with 0.87 nm lead width. The selection of hexagonal design was based on its similarity to fabricated nanostructures of two-dimensional honeycomb materials including graphene and transition-metal dichalcogenides which commonly appear as hexagonal flakes or quantum dots QDs \cite{Chen2019-gb}\cite{Da_Costa2016-os} \cite{Lee2015-sq}. The chosen geometry enables researchers to examine edge termination effects on transport through direct comparison with existing studies. The hexagonal design contains more defined edge regions than disc shapes which increases the impact of edge-related states on electronic and spin transport properties. The configuration provides an ideal setting to study how zigzag and armchair boundaries control confinement effects and spin polarization in borophene nanostructures.

The band structure of the zigzag-edged hexagons and the armchair-edged hexagons exhibits distinct behaviors as shown in Fig.\ref{hexagon_T}: the zigzag-edged structure lacks a bandgap while the armchair-edged structure maintains one. The zigzag structure within this geometry no longer displays its typical zero transmission plateau compared to the disc geometry, although the zero points remains evident in Fig.~\ref{hexagon_T}(a). The transport properties of the hexagonal structure strongly depend on the edge states according to this observation. The hexagonal geometry possesses sharper terminations than the disc geometry which leads to stronger localization of edge modes. The stronger edge effects prevent the development of widespread plateaus but allow zero-energy states to remain localized. The armchair-edged hexagon displays transport characteristics identical to the armchair disc structure by showing a distinct plateau around the Fermi energy Fig.~\ref{hexagon_T}(b) which demonstrates its semiconducting nature. The comparison demonstrates that edge geometry sharpness leads to stronger edge-state contributions which change transport signatures especially in systems with zigzag-terminated edges.
\begin{figure}
    \centering
    \includegraphics[width=1\linewidth]{hexagon_T.pdf}
    \caption{Band structure (left) and charge transmis-
sion (right) for borophene hexagonal systems with edge
orientations of (a) zigzag and (b) armchair.}
    \label{hexagon_T}
\end{figure}

\begin{figure}[h]
    \centering
    \includegraphics[width=1\linewidth]{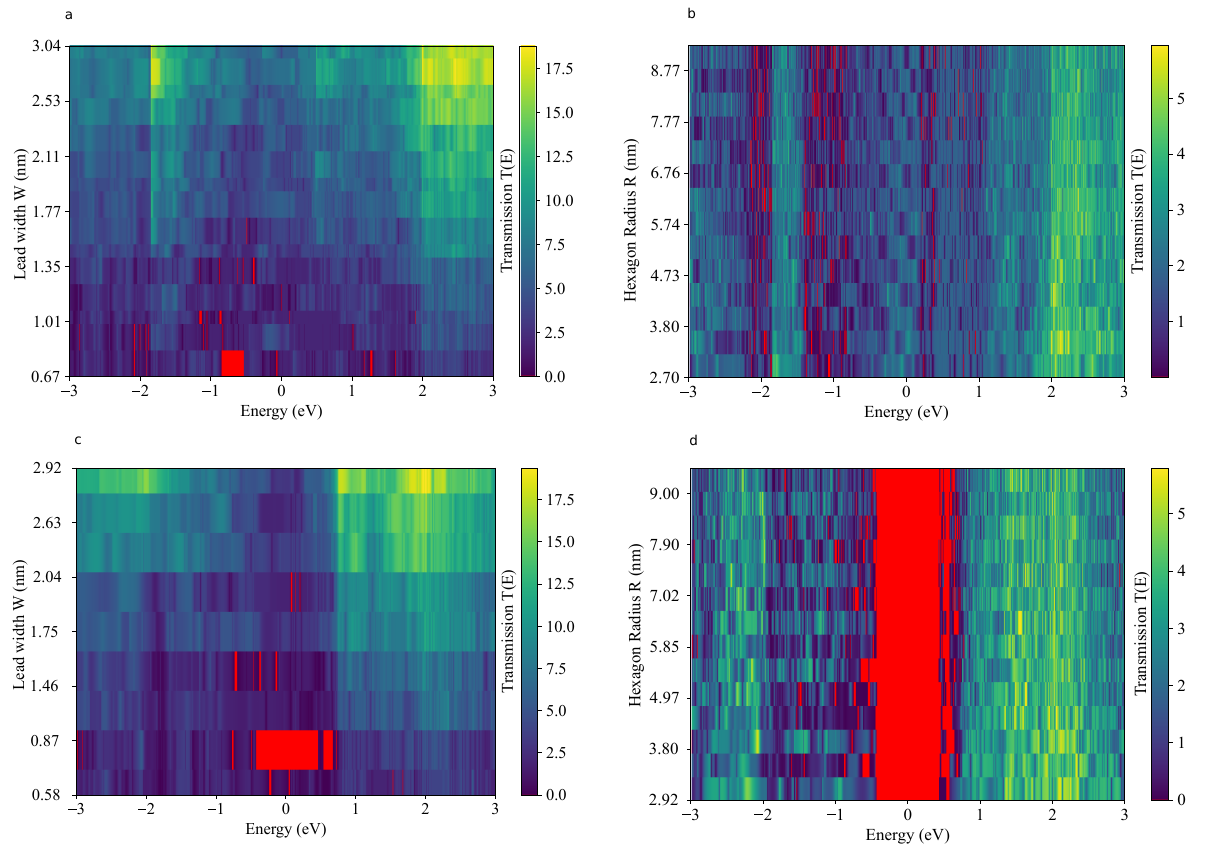}
    \caption{Transmission of initial desired hexagon with zigzag edges (upper row) and armchair edges (bottom row) by changing leads width (a, c) and disc radius (b, d). Red color indicate zero transmission.}
    \label{heat_h}
\end{figure}

Figure~\ref{heat_h} illustrates the dependence of the transmission spectra on the width of the leads and the radius of the hexagonal region. The increase in lead width activates more transport channels because wider leads enable additional propagating modes to participate in conduction. The transmission plateau shows clear definition for zigzag-edge leads when the width stays below 0.84 nm Fig.~\ref{heat_h}(a) which indicates quantum confinement dominates this regime along with limited mode availability. However, even when the radius of the hexagon is varied for this initial width, no plateau reappears Fig.~\ref{heat_h}(b), 
The sharp edges and corners of the hexagon create an interference with the translational symmetry, which leads to strong reflections of electronic wavefunctions that generate localized edge states. The multiple edges create two opposing quantum interference effects at the boundaries. These block extended conduction paths and reduce transmission features even though the band structure remains metallic\cite{Romanovsky2012-fs} \cite{Konopka2015-tz}.
The armchair-edge configuration displays distinct zero-transmission areas when the lead widths fall between 0.72 nm and 0.9 nm Fig.~\ref{heat_h}(c). The armchair edge semiconducting properties along with its stronger confinement at narrow widths lead to these zero-transmission regions. The hexagon’s radius variation produces minimal changes in the zero-conduction region locations similar to the disc geometry Fig.~\ref{heat_h}(d). Quantum interference and lead-edge confinement control the transport suppression instead of the central hexagonal region's size.

\begin{figure}[h]
    \centering
    \includegraphics[width=1\linewidth]{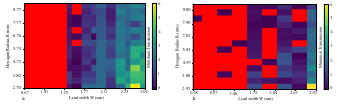}
    \caption{Minimum transmission (red color) per vary range of lead width and disc radius with zigzag (a) and armchair (b) edges.}
    \label{heat_r_w_h}
\end{figure}
As shown in Fig.~\ref{heat_r_w_h}(a), the transmission behavior of the hexagonal borophene device remains largely unaffected by increasing the radius. When the leads with zigzag edges are narrower than roughly 1.7 nm, the system exhibits complete suppression of transmission, reflecting strong confinement and the absence of available conducting channels. On the other hand, for armchair-edged leads Fig.\ref{heat_r_w_h}, enlarging the width gradually removes the zero-transmission regions, indicating that the system becomes more delocalized and allows charge carriers to propagate more freely.
\subsubsection{Spin Transport of Hexagon Geometry}
\begin{figure}[h]
    \centering
    \includegraphics[width=1\linewidth]{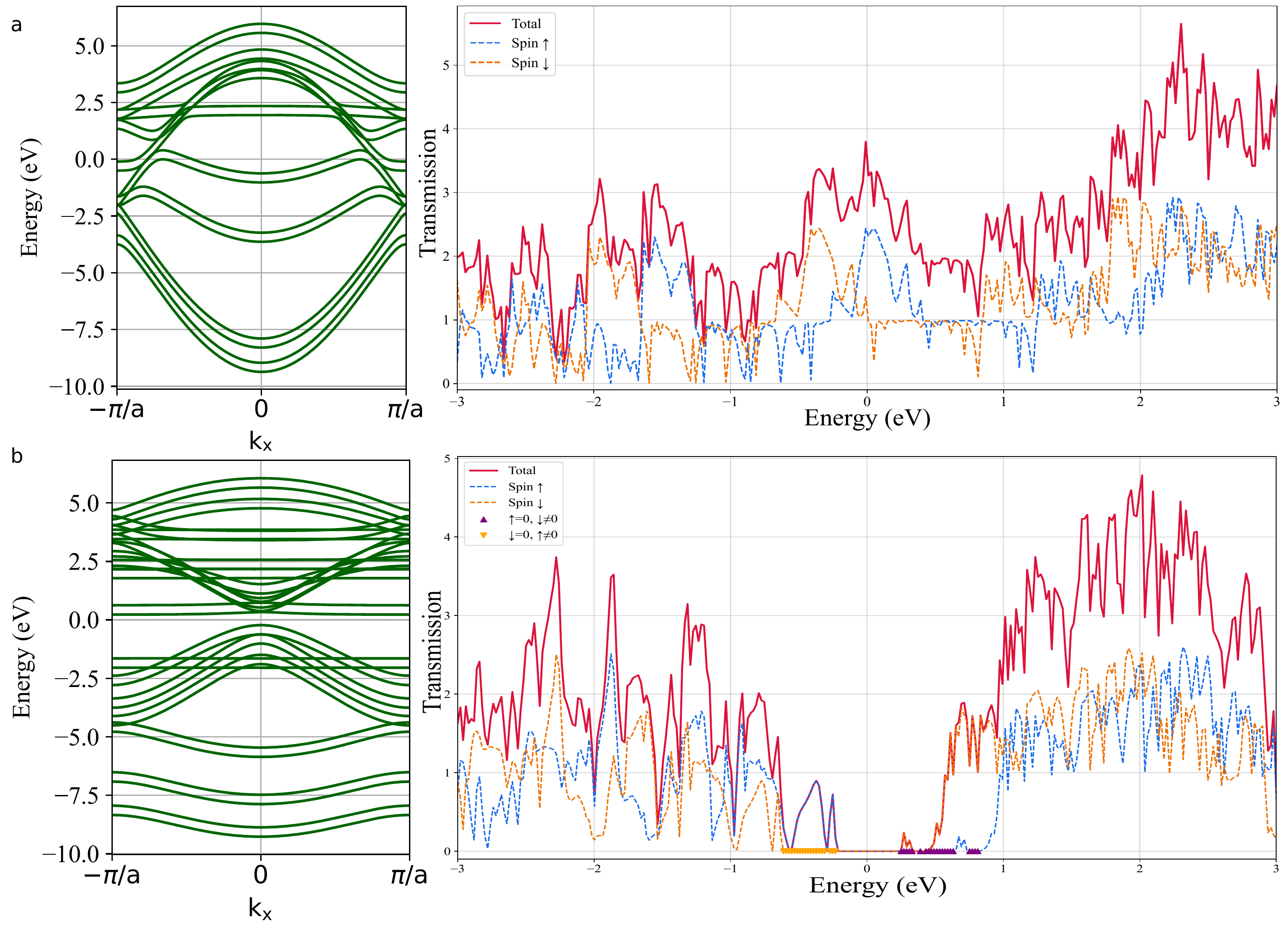}
    \caption{Band structure (left ) and spin-resolved transmission (right ) for borophene hexagonal systems with edge orientations of (a) zigzag and (b) armchair.}
    \label{T_m_h}
\end{figure}
\begin{figure}[h]
    \centering
    \includegraphics[width=1\linewidth]{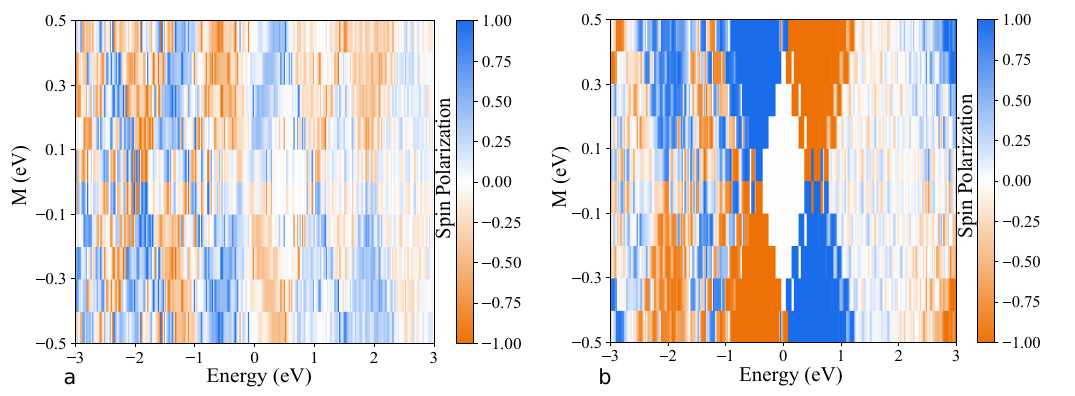}
    \caption{Spin polarization for vary exchange fields for hexagon geometry with zigzag (a) and armchair (b) edges as function of the energy of incoming electrons.}
    \label{H_m_h}
\end{figure}

By applying a nonlocal exchange field of M=0.2eV, the band structures of the hexagonal systems become spin-split, and corresponding modifications appear in their transmission spectra, as shown in Fig.~\ref{T_m_h}. For the zigzag-edged configuration Fig.~\ref{T_m_h}(a), full spin polarization is not achieved within the studied energy window, and even varying the exchange field strength fails to induce complete spin-selective transport Fig.~\ref{H_m_h}(a). In contrast, the armchair-edged system Fig.~\ref{T_m_h}(b) displays distinct regions of full spin polarization: fully spin-up transport occurs between $−0.69$ eV and $−0.23$ eV, while fully spin-down transport appears in the ranges $0.23e$V to $0.65$eV and
$0.75$ eV to $0.83$ eV. Similarly to the disc geometry, the hexagonal configuration with armchair edges orientation supports well-defined spin-polarized transport across a broad energy range, and this behavior remains robust against moderate variations in the exchange field.The difference between the two edge types results from their fundamental electronic properties. The metallic-like states of zigzag edges create high spin degeneracy which produces incomplete spin filtering when exchange fields are moderate. The natural semiconducting properties of armchair edges enable the exchange interaction to create separate spin-dependent energy gaps which result in complete spin polarization in selected energy regions. Other two-dimensional materials with honeycomb lattices including graphene and silicene nanoribbons show edge-dependent spin-selective behavior because the exchange interaction and geometry determine their spin transport properties \cite{Rahmani-Ivriq2020-uz} \cite{Shakouri2015-lt}\cite{Wimmer2008-uv}.

\section{Conclusion}
The research investigates how quantum dot confinement shapes and contact lead arrangements along with edge designs determine charge and spin transport properties in $\beta 12$ borophene quantum dots through tight-binding and NEGF methods. The methodical analysis of disc and hexagon central areas as well as zigzag and armchair lead terminations demonstrates consistent patterns in charge-transmission spectra and spin-resolved transport when exposed to a proximity-induced exchange field.

Charge transport results demonstrate an edge-based distinction between systems that connect to zigzag leads and those that connect to armchair leads. Systems connected to zigzag leads exhibit metallic band structures with transmission fluctuations and zero-transmission points whereas armchair-connected systems feature flat bands near the Fermi level and transmission plateaus consistent with semiconducting behavior. The observed distinctions remain stable across both disc and hexagon central geometries and can be seen in the band-structure and transmission plots shown.

The geometry of the lead primarily its width significantly controls the transport characteristics because narrow leads increase both quantum confinement and destructive interference which leads to permanent zero-transmission areas but broader leads provide more propagating channels that reduce these transport blockades. Our research demonstrates specific lead width values that mark the end of zero-transmission behavior at approximately 1.01 nm for zigzag leads and 0.87 nm for armchair leads in our studied parameter sets. The energy positions of these suppression features remain nearly unaffected by changes in the central-region radius because lead-edge confinement proves stronger than central-area size in determining the observed transmission gaps.
Time-reversal symmetry breaking with a nonlocal exchange field causes the leads' bands to develop distinct spin splitting which creates specific energy windows of complete spin polarization. Zigzag-connected discs produce small fully spin-polarized energy intervals (such as full spin-up between −1.07 eV and −0.97 eV and full spin-down between −0.67 eV and −0.57 eV) whereas armchair-connected discs generate much larger complete spin-polarization energy ranges (such as full spin-up between −0.61 eV and −0.23 eV alongside corresponding spin-down intervals). These findings confirm that the interaction between confinement effects and edge termination methods enables energy-selective spin filtering capabilities.

The hexagonal core region shows different behavior patterns from disc geometry because zigzag-edged hexagons do not achieve total spin polarization within the tested energy range regardless of exchange field strength but armchair-edged hexagons display multiple distinct full-polarization energy intervals including spin-up from approximately −0.69 eV to −0.23 eV and several spin-down windows at positive energies). The hexagonal terminations are more defined than disc terminations so they enhance edge state localization and alter transmission plateau formation and zero-energy state appearance yet armchair terminations continue to provide superior operational ranges for spin-selective transport.

The research findings demonstrate two essential practical insights. The edge treatment of borophene-based nanodevices acts as an effective control mechanism to modify both charge conduction characteristics between metallic and semiconducting-like behavior and spin filtering operational energy ranges. The design of functional nanoscale spintronic elements requires equal attention to lead dimensions and lead matching conditions as it does to central scatterer geometry. The armchair-connected structures (both disc and hexagon) show potential as adaptable spin filters with energy selectivity which remain stable against small changes in exchange strength.

\bibliographystyle{elsarticle-num}
\bibliography{ref}
\end{document}